# Tuning the Magnetic and Electronic Properties of Strontium Titanate by Carbon Doping

Hui Zeng[1], Meng Wu[1], Hui-Qiong Wang[1,2*], Jin-Cheng Zheng[1,2*], Junyong Kang[1]

[1] Fujian Provincial Key Laboratory of Semiconductors and Applications, Collaborative Innovation Center for Optoelectronic Semiconductors and Efficient Devices, Department of Physics, Xiamen University, Xiamen 361005, P. R. China

[2] Xiamen University Malaysia, Sepang 43900, Selangor, Malaysia

*Corresponding authors. Emails: hqwang@xmu.edu.cn; jczheng@xmu.edu.cn

## Abstract

The magnetic and electronic properties of strontium titanate with different carbon dopant configurations are explored using first-principles calculations with a generalized gradient approximation (GGA) and the GGA+U approach. Our results show that the structural stability, electronic properties and magnetic properties of C-doped $SrTiO_3$ strongly depend on the distance between carbon dopants. In both GGA and GGA+U calculations, the doping structure is mostly stable with a nonmagnetic feature when the carbon dopants are nearest neighbors, which can be ascribed to the formation of a C-C dimer pair accompanied by stronger C-C and weaker C-Ti hybridizations as the C-C distance becomes smaller. As the C-C distance increases, C-doped $SrTiO_3$ changes from an n-type nonmagnetic metal to ferromagnetic/antiferromagnetic half-metal and to an antiferromagnetic/ferromagnetic semiconductor in GGA calculations, while it changes from a nonmagnetic semiconductor to ferromagnetic half-metal and to an antiferromagnetic semiconductor using the GGA+U method. Our work demonstrates the possibility of tailoring the magnetic and electronic properties of C-doped $SrTiO_3$, which might provide some guidance to extend the applications of strontium titanate as a magnetic or optoelectronic material.

1. Introduction

As one of the most fascinating electronic ceramic materials, strontium titanate ($SrTiO_3$) has attracted much attention because of its high static dielectric constant [1, 2], large electron effective mass and large Seebeck coefficient [3], which has led to promising technological applications [4-7]. Due to its high efficiency, stability and strong reducing ability in the photocatalytic field, $SrTiO_3$ can also be employed as a feasible photocatalyst [8-13].

It has been found that the performance of $SrTiO_3$ can be dramatically modified through doping [14-18]; for example, doping is chosen as a typical approach to reduce the bandgap of $SrTiO_3$ to enhance its response towards the solar spectrum. The absorption of visible light in N-doped $SrTiO_3$ can be attributed to the induced isolated states above the valence band maximum (VBM) [19, 20]. Doping with only Rh to substitute for Sr can significantly reduce the photoabsorption energy by introducing localized acceptor states. Codoping with La and Rh leads to the formation of clean band structures rather than mid-gap states [8]. Doping approaches mainly include (i) anion doping [5, 21], (ii) cation doping [14, 22, 23], (iii) cation-anion codoping [13, 24-26], (iv) anion-anion codoping [27], and (v) cation-cation codoping [5, 28]. In particular, the anion dopant of carbon (C) to replace oxygen has attracted much attention and is considered one of the appropriate anion alloy elements to introduce acceptor levels above the VBM of SrTiO$_3$ because its size is closer to the size of the substituted host oxygen atoms, which is favorable for effective incorporation. Theoretical results indicate that C-doping induces a small change in the band gap, and the new absorption edge in the visible light region is attributed to the isolated impurity band [19, 29]. For codoping research, the C element is also considered one of the best candidates for generating the original impurity level between the valence band and conductor band [20]. Investigation of SrTiO$_{2.875}$C$_{0.125}$ has revealed that it is a magnetic semiconductor, with a magnetic moment of approximately 2.00 $\mu_B$ [30]. However, the effect of coupling between C atoms on structural stability, magnetic moment and electronic structure remains unclear; for instance, whether the induced carbons tend to align in a chain sandwiched with Ti atoms or prefer to locate far away from each other remains unclear. For the purpose of exploring the magnetic moment distributions and their related electronic structures in detail, we employ density functional theory (DFT) to simulate seven different configurations separated by different C-C distances in SrTiO$_{2.75}$C$_{0.25}$. The results of generalized gradient approximation (GGA) and GGA+U are calculated for comparison. We hope the study of carbon coupling in SrTiO$_{2.75}$C$_{0.25}$ systems can provide theoretical guidance for engineering doping approaches and designing new materials.

2. Computational details

Our first-principles calculations are performed using the Vienna ab initio Simulation Package (VASP) [31, 32] based on DFT [33] with projected augmented wave (PAW) potentials. The generalized gradient approximation (GGA) parameterized by Perdew-Burke-Ernzerhof (PBE) is used for the exchange-correlation functional [34]. We employ an on-site U = 5.8 eV on the Ti 3d orbitals to consider the strongly correlated effects and provide a proper description of the Ti 3d localized states [22, 35-37]. For some typical configuration cases, we also investigate the magnetic properties as a function of U to examine the localization influence. The atomic reference electron configurations for the PAW potentials are $4s^24p^65s^2$ for Sr, $3s^23p^63d^24s^2$ for Ti, $2s^22p^4$ for O and $2s^22p^2$ for C.

For the calculations of bulk $SrTiO_3$ and different configurations sets for the deficient $SrTiO_3$ systems, a large 2×2×2 supercell containing 40 atoms is chosen. In the supercell, two O atoms are replaced by two C atoms, namely, $SrTiO_{2.75}C_{0.25}$. The doped model is shown in Fig. 1, in which the position of carbon atom i (location of first dopant in supercell) remains invariable, while the other carbons (second dopant in the supercell) are labeled A-G away from i with increasing separation. Hereinafter, we name A-G carbon-doped combinations configurations A-G. The effects of the separating distance between dopants on properties were investigated in detail. Electronic structures are calculated on the corresponding optimized crystal geometries. A cutoff energy of 450 eV for the plane-wave basis set and a 5×5×5 k-point grid centered at the Gamma (or Γ) point are used. The convergence threshold for self-consistent iteration is set at $10^{-5}$ eV, and all the atomic positions are fully optimized until all components of the residual forces are smaller than 0.01 eV/Å.

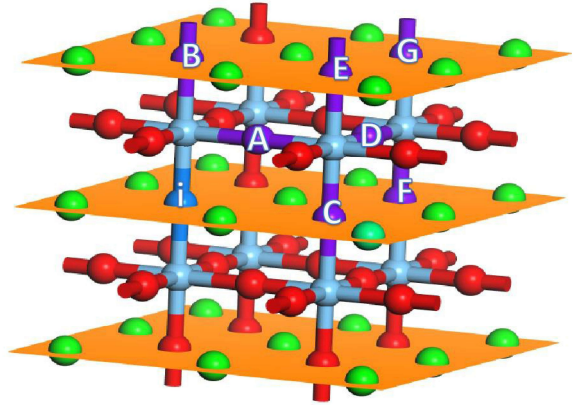

**Fig. 1.** The model illustrates different A-G configurations in SrTiO$_{2.75}$C$_{0.25}$ with increasing separation from $\frac{\sqrt{2}}{2}a$ to $\sqrt{3}a$ ($a$ is the lattice constant of cubic SrTiO$_3$). Color codes: green for Sr, red for O atoms, violet red for Ti, royal blue i for the fixed C dopant and blue violet A-G for the possible positions of the second C dopant.

## 3. Results and discussion

### 3.1 Magnetic property

To explore the spin-polarized properties, we discuss the specified magnetic moment distributions in seven different SrTiO$_{2.75}$C$_{0.25}$ configurations. Detailed calculation results are shown in Table 1. To explain the most stable magnetic states with carbon impurities incorporated within the accepted lattice, we employ the relative energy ($\Delta E$) in different SrTiO$_{2.75}$C$_{0.25}$ configurations based on the following equation: $\Delta E = E_{meta} - E_{stable}$, where $E_{meta}$ is the total energy of the calculated metastable configuration, and $E_{stable}$ is the total energy of the stable configuration that has the lowest energy among different magnetic states in C-doped $SrTiO_3$ (eV/dopant), which is referred to as the NM (nonmagnetic) state in configuration A ($E_{stable}$ is set as 0 eV). The initial magnetic calculations under ferromagnetic (FM) and antiferromagnetic (AFM) alignments for configuration A both converged to NM states, indicating an NM nature for the ground magnetic state (GMS). With respect to all other configurations, configuration B is endowed with the highest total energy and illustrates the most unstable configuration, which is consistent with the following investigation of electronic properties. The relative energy of other configurations in FM/AFM magnetic states and NM states is approximately 2.5 eV/2.9 eV higher than that of configuration A, respectively. In addition, compared to pristine $SrTiO_3$, configurations A-G are characterized by approximately 3.25 eV, 8.43 eV, 8.24 eV, 8.08 eV, 8,24 eV, 8.33 eV, and 8.27 eV higher energies, respectively.

As the dopant-dopant distance increases, the GMS for various configurations switches to FM or AFM magnetic states. All the magnetic configurations are endowed with similar FM magnetic moments of approximately 4.0 $\mu_B$, which can be qualitatively ascribed to the magnetic moment of the carbon dopants. Bannikov *et al.* revealed that the overall magnetic moments in $SrTiO_{2.875}C_{0.125}$ (with one carbon dopant) acquired a net spin moment of 2.02 $\mu_B$ [30], and the magnetic moment originated mainly from the spin splitting of states for impurity carbon atoms, which is consistent with our results. In fact, the unpaired or uncompensated electrons induced by carbon dopants lead to the appearance of magnetism, which may provide additional approaches to yield magnetic moments by introducing nonmagnetic ions.

The relative contributions from carbon atoms to the total magnetization in different arrangements are summarized in Table 1, which manifests the following three features: (1) No noticeable magnetic moments are observed for carbon atoms in configuration A; (2) Compared within magnetic systems, configuration B depicts a rather small total magnetic moment, and the magnetic moments on two isolated carbon atoms are unequal; (3) There appears to be an equal contribution of approximately 0.7 $\mu_B$ for the magnetic moment per carbon atom in configurations C-G. Table T1 in the supplemental materials shows the averaged contribution of the Sr atom, Ti atom, and O atom to magnetism. It shows that no noticeable contribution of magnetic moments is related to Sr and Ti atoms, while relatively small localized magnetic moments appear at O atoms, suggesting that the resolved magnetic moment originates mainly from impurity carbon atoms. We note that due to the default radius set for the integration of the magnetic moment in VASP, the total magnetic moment is less than the sum of the magnetic moment of each atom, which does not affect our conclusion, especially for the trend we will discuss in the following sections. Configuration A shows a more favorable NM GMS even though it is associated with unpaired electrons similar to configurations B-G. Thus, to explore the coupling mechanism of the observed peculiarity, we will present the investigation of NM in configuration A later on.

Table 1. Lists of the related energies (eV/dopant) in nonmagnetic (NM), ferromagnetic (FM) and antiferromagnetic (AFM) states for the 2x2x2 supercell, total magnetic moment (MM$_{tot}$) and localized magnetic moment on carbon (MM$_C$) in units of $\mu_B$, nature of the ground magnetic state (GMS) for different configurations in SrTiO$_{2.75}$C$_{0.25}$. The lowest total energy of configuration A is taken as a reference. The data in bold font highlight the lowest energy (the most stable state) among the NM, FM and AFM alignments for each configuration (note, if the energy difference between FM and AFM is relatively small, e.g., less than typical 25

meV, then both FM and AFM are labeled possible GMSs to take into account the temperature effects and numerical uncertainty).

| Configurations | ΔE(NM/FM/AFM) | MM$_{tot}$(FM/AFM) | MM$_C$(FM/AFM) | GMS |
|---|---|---|---|---|
| A | 0 | 0 | 0 | NM |
| B | 2.856/**2.549**/2.564 | 2.871/1.257 | 0.278, 0.737/0.750, -0.228 | FM/AFM |
| C | 2.829/2.493/**2.470** | 3.982/-0.001 | 0.676, 0.676/0.643, -0.643 | AFM/FM |
| D | 2.902/**2.420**/2.421 | 4.001/0.005 | 0.681, 0.681/0.681, -0.681 | FM/AFM |
| E | 2.765/2.493/**2.430** | 3.726/-0.002 | 0.646, 0.646/0.623, -0.623 | AFM |
| F | 2.810/2.537/**2.470** | 3.994/-0.002 | 0.669, 0.669/0.636, -0.636 | AFM |
| G | 2.996/2.508/**2.505** | 3.997/-0.002 | 0.661, 0.661/0.661, -0.661 | AFM/FM |

## 3.2 Structural stability

To further acknowledge the differences in the doped configurations, we investigate the structural stability and charge transfer information. The relaxed lattice parameters for pristine SrTiO$_3$ are $a=b=c=3.935$ Å, which is in good agreement with the experimentally obtained lattice constants ($a=b=c=3.905$ Å). The optimized C-C distance in different combinations ranges from 1.297 Å to 6.868 Å (see supplemental materials Fig. S1), and the interrelated calculated results are summarized in Table 2.

Table 2. Calculated distances between the C atom and the nearest Sr, Ti, and C atom (Å) in SrTiO$_{2.75}$C$_{0.25}$ for seven different configurations, where Δ represents the changes in the bond length before and after relaxation. Negative and positive values represent attraction and repulsion, respectively. R$_d$ (Å) corresponds to the distance between two nearest carbon dopants before relaxation, where $a$ is the lattice constant of pristine SrTiO$_3$. Parameters $\Delta a$, $\Delta b$ and $\Delta c$ represent the variation in the lattice constants (Å) after relaxation in different configurations versus relaxed pristine SrTiO$_3$ lattice parameters, respectively.

| Configuration | R$_d$ | Δ(C-C) | Δ(C-Sr) | Δ(C-Ti) | $\Delta a$ | $\Delta b$ | $\Delta c$ |
|---|---|---|---|---|---|---|---|
| A | $\sqrt{2}a/2$ | -1.496 | -0.141 | 0.216 | 0.025 | 0.019 | 0.025 |
| B | $a$ | 0.039 | 0.042 | -0.042 | 0.009 | 0.009 | 0.053 |
| C | $a$ | -0.009 | 0.044 | 0.084 | 0.005 | 0.004 | 0.080 |
| D | $\sqrt{6}a/2$ | 0.076 | 0.055 | 0.105 | 0.042 | 0.014 | 0.042 |
| E | $\sqrt{2}a$ | 0.037 | 0.044 | 0.089 | 0.003 | 0.005 | 0.078 |
| F | $\sqrt{2}a$ | -0.017 | -0.008 | 0.096 | 0.002 | 0.002 | 0.082 |
| G | $\sqrt{3}a$ | 0.028 | 0.032 | 0.083 | 0.004 | 0.004 | 0.081 |

The substantial displacement of the C-C bond length before and after a relaxation of approximately 1.5 Å in configuration A can be attributed to the strong attraction between carbon atoms inducing a dimer pair with a C-C distance of 1.297 Å. The dimer pair between carbon and carbon atoms has also been reported in monolayer ZnO when two neighboring O atoms are substituted by C atoms [38], resulting in an n-type semiconductor with a NM ground state. In addition, the significant displacement of the C-C bond length is responsible for the considerable shift of C-Sr and C-Ti, whereas the attraction or repulsion in all other configurations is less than 0.1 Å. Configurations C and F also show a tiny C-C attraction and only the two configurations on the Sr-O face, while other configurations except for configuration A express repulsion between carbons. The altered distances for C-Sr show similar results accompanied by a repulsion of approximately 0.04 Å in all configurations except for configurations A and F. Ti atoms tend to attract the nearest carbon atom in configuration B, whereas Ti atoms move away from carbon atoms for all other cases.

To further investigate the structural properties in more detail, we examine the interactions between C atoms and the nearest Sr and Ti ions, as well as the effects of such interactions on volume or lattice parameters. Herein, for instance, we select configuration A with the nearest neighbor C-C and configuration B with the second nearest neighbor C-C to perform the comparison. The optimized structures of configuration A and configuration B are presented in Fig. 2(a)(c). First, as shown in Table 2, compared to pristine $SrTiO_3$, C-doped SrTiO3 shows a larger volume (all positive $\Delta a$, $\Delta b$ and $\Delta c$) in all configurations. The relaxed lattice parameters in configuration A are $a=c=3.960$ Å and $b=3.954$ Å, which are approximately 0.02 Å larger than those of pristine $SrTiO_3$. As mentioned above, two carbons lean very closer and tend to form a C-C dimer pair in configuration A. The changes in the bond length before and after relaxation for C-C, C-Sr, and C-Ti are 1.496 Å, 0.141 Å, and 0.216 Å, respectively. For configuration B, as marked in Fig. 2(c), two doped carbons align linearly along the C-Ti-C axis. The $C_1$-$Ti_1$ bond length is decreased by 0.042 Å, while the $C_2$-$Ti_1$ bond length is increased by 0.081 Å. This triggers a small rotation of the $O_1$-$Ti_1$-$O_2$ angle and significantly increases the $c$ lattice constant by approximately 0.053 Å, while the lattice constants $a$ and $b$ remain unchanged.

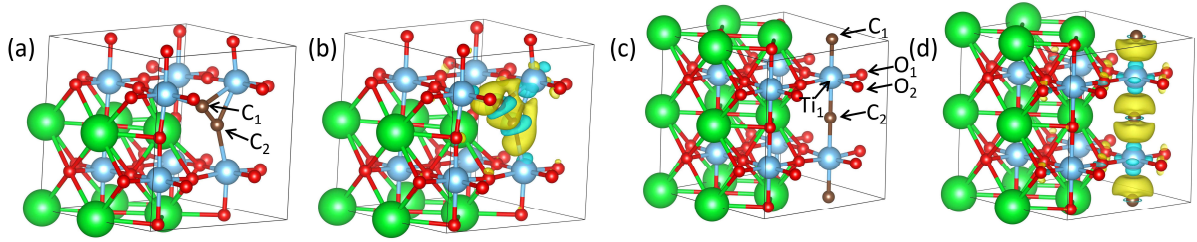

**Fig. 2.** Panels (a)(c) show the relaxed structures of configurations A and B, respectively. Panels (b) and (d) represent the corresponding charge density difference plots. The isosurface level value is set as 0.005 e/Å³, and the yellow and blue violet areas represent charge accumulation and charge depletion, respectively.

To understand the charge transform and redistribution, we employ the following formula to calculate the charge density difference $\rho = \rho_{total} - \rho_{STO} - \rho_{C_1} - \rho_{C_2}$, where $\rho_{total}$, $\rho_{STO}$, $\rho_{C_1}$ and $\rho_{C_2}$ are the charge of the $SrTiO_{2.75}C_{0.25}$ system, pristine $SrTiO_3$, carbon1 and carbon2 atoms, respectively. The charge density difference plots of configuration A and configuration B are shown in Fig. 2(b) and (d), respectively. Herein, the isosurface level value is set as 0.005 e/Å³, and the yellow or red areas represent charge accumulation, while blue violet areas represent charge depletion. The variation in the electronic distribution due to charge transfer to the carbon is rather obvious. Significant charge accumulation near the carbon and a small amount of charge depletion around the titanium atoms closest to the carbon in configurations A and B gives rise to strong electrostatic interactions between titanium atoms and carbons. Other charge difference plots for configuration C-G are characterized with similar properties and are listed in supporting information S2. In addition, Bader charge analysis indicates that each C dopant acquires on average approximately 0.75e charge in configuration A and 0.91e charge in configuration B.

As mentioned above, the relative energy (ΔE) is important to evaluate the coupling strength of the substitutional dopant. Figure 3(a) illustrates the calculated relative energy in different configurations, and it can be seen that the formation of a C-C dimer pair in configuration A is the most favorable with the lowest energy, where two impurity carbons tend to strongly attract each other. Moreover, the energy is far lower than that for the other six configurations. Except for configuration A, a comparatively low energy is observed in configuration D. Two doped carbons aligned linearly along the C-Ti-C axis in configuration B are endowed with a relatively high energy, which suggests that it may be difficult to synthesize configuration B through carbon doping. Figure 3(b) illustrates the calculated relative energy versus different separated distances between two carbon atoms (before relaxation) in the $SrTiO_{2.75}C_{0.25}$ system. Even with the same C-C distance, the total energy depicts visible diversity for configurations B and C, as well as E and F. It is found that the C-C pair tends to remain at the Sr-O face. For example, compared to configuration B, the C-C pair tends to remain at the Sr-O face in configuration C, in which the pairs align along the x-axis with no other atoms between them. The C-C pair also prefers to remain at the Sr-O face in configuration F with respect to configuration E.

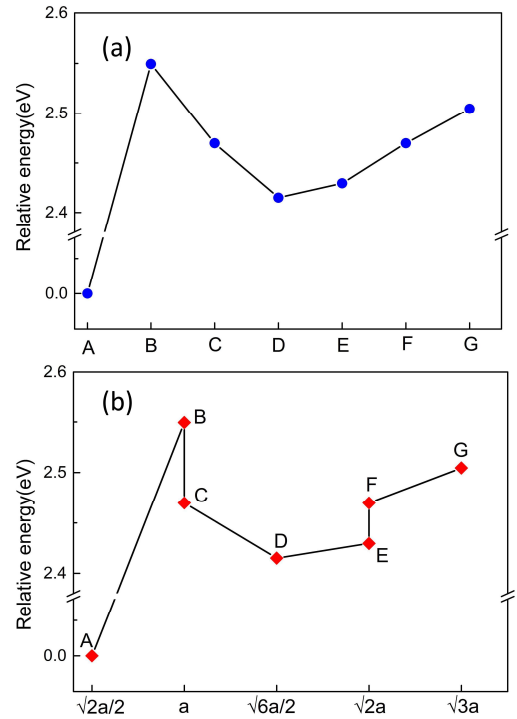

**Fig. 3.** Panel (a): Calculated relative energy (eV/dopant) of different configurations (the model of atomic configurations is shown in Fig. 1). Panel (b): Separated distance (before relaxation) between two carbon atoms in the SrTiO$_{2.75}$C$_{0.25}$ system. The lowest energy in configuration A is set as a reference.

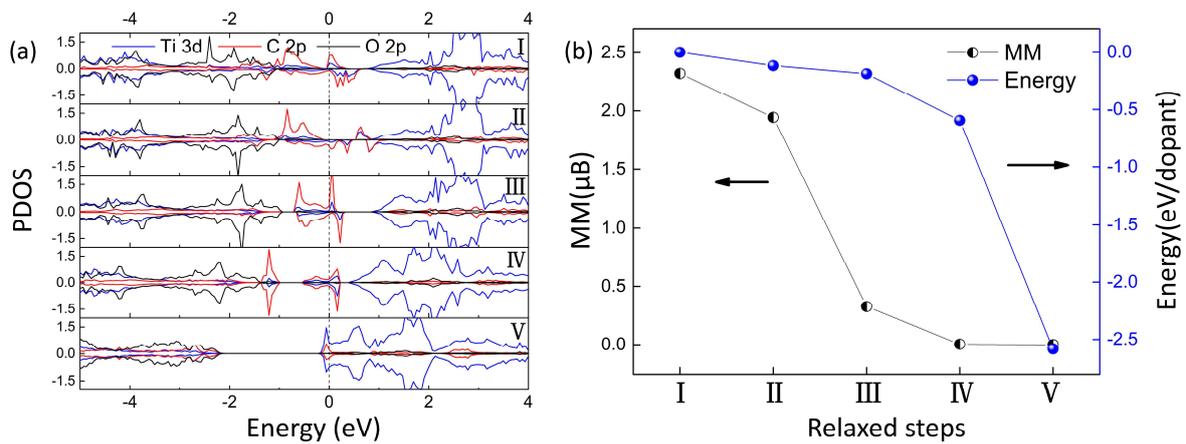

**Fig. 4.** (a): The partial density of states (PDOS) of configuration A (two-carbon-doped SrTiO$_3$) in different relaxation steps (from initial ideal oxygen sites to final relaxed positions). The detailed descriptions from steps

I to V are reported in the main text, and the model of configuration A is displayed in Fig. 1. Panel (b) shows the corresponding total magnetic moments ($\mu_B$) and energies (eV/dopant) versus different relaxed steps.

To provide insights into the origin of the nonmagnetic property in configuration A and to explain the induced different magnetic features in other doped configurations, we further analyze the partial density of states (PDOS) of Ti 3d orbitals, C 2p orbitals and O 2p orbitals with different relaxed steps, as shown in Fig. 4(a). Three criteria are applied in our calculations: (1) The C-C bond length before relaxation is 2.792 Å, named step I (initial step, ideal doping site), while, after relaxation, it is 1.296 Å, labeled step V (final step, completely relaxed site); (2) The coordinates for the two carbon atoms at steps I and V are (0.250, 0.500, 0.250), (0.500, 0.750, 0.250) and (0.346, 0.537, 0.250), (0.463, 0.654, 0.250), respectively. Other intermediate steps indicate that the coordinates for the two carbon atoms sit between I and V. For instance, step III is associated with the carbon coordinates of (0.298, 0.519, 0.250) and (0.482, 0.702, 0.250), which correspond to the middle position of the separated C-C distance between the initial step I and the final step V in configuration A; (3) coordinates of the carbon atoms remain unchanged under every condition in our calculations.

As illustrated in Fig. 4(a), at the initial step I, C 2p and Ti 3d orbitals show a strong $pd$ orbital hybridization near the Fermi level, and most of the unequal spin-up and spin-down electrons are ascribed to the foreigner carbons, which introduce a remarkable magnetic moment. As the C-C distance approaches step V, the hybridization between C 2p and Ti 3d gradually weakens. Simultaneously, Ti 3d orbitals are pulled down to lower energies and depict a more dominant contribution with respect to equal spin-up and spin-down channels, which results in a decrease in the magnetic moment, as shown in Fig. 4(b). The magnetic moment varies from 2.3 to 0 $\mu_B$ when the step evolves from I to V. Therefore, the nonmagnetic character mainly originates from the competition between C-C hybridization and C-Ti hybridization, where the stronger C-C hybridization and weak C-Ti hybridization give rise to equal spin-up and spin-down channels. Figure 4(b) also shows the corresponding energies with respect to different relaxed steps. The energy in step I is set as the reference energy (i.e., set as 0). An obvious decrease in energies from step I to step V indicates a more stable state as the two carbon atoms get closer, which is consistent with our research below.

### 3.3 Electronic structure

The calculated total density of states (DOS) and the PDOS for the O 2p, Ti 3d $t_{2g}$ and Ti 3d $e_g$ orbitals for pristine $SrTiO_3$ are shown in Fig. 5(a) and 5(b), respectively. This illustrates that pristine $SrTiO_3$ possesses a band gap of 1.69 eV [39-41], which is in good agreement with other calculations. This value is smaller than the experimental value of 3.20 eV due to the well-known underestimation of the band gap by LDA or GGA. As seen from Fig. 5(b), the valence bands of pristine $SrTiO_3$ are predominantly composed of O 2p orbitals, and the conduction bands mainly correspond to Ti 3d orbitals. Because of the cubic crystal field in the $TiO_6$ octahedron surrounded by six nearest-neighbor O atoms, the empty Ti 3d orbitals split into three lower energy $t_{2g}$ orbitals ($3d_{xy}$, $3d_{yz}$, $3d_{zx}$) and two higher energy $e_g$ orbitals ($3d_{z^2}$, $3d_{x^2-y^2}$). Pristine $SrTiO_3$ exhibits a nonmagnetic property due to the symmetric populations of spin-up and spin-down channels.

The calculated spin-polarized DOS for configurations A-G are shown in Fig. 6(a)-(g). Compared to pristine $SrTiO_3$, the introduced carbon impurities lead to the appearance of a new impurity 2p band localized in the band gap of the matrix and shift the Fermi level in all configurations. In addition, the splitting of the impurity band into spin-resolved bands and the unequal spin-up and spin-down electrons in configuration B-G contribute to a remarkable magnetic moment, while equal spin-up and spin-down channels result in the nonmagnetic ground state in configuration A. Based on electronic structures and magnetic features, three different $SrTiO_{2.75}C_{0.25}$ are classified. The C-doped $SrTiO_3$ undergoes a change from n-type NM metal to FM/AFM half-metal and to AFM/FM semiconductor with increasing dopant-dopant distance. Detailed electronic structures for different categories will be presented in the following subsections.

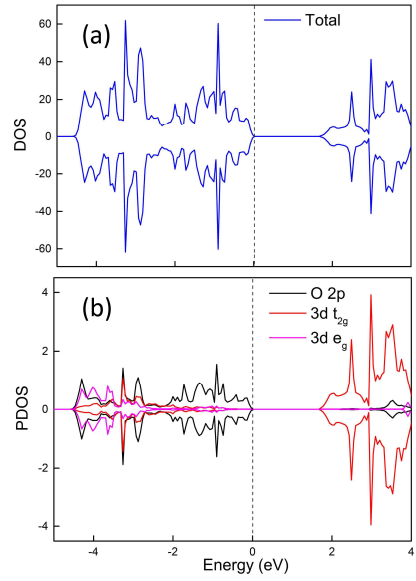

**Fig. 5.** (a) Calculated total density of states (DOS) for pristine SrTiO$_3$ and (b) the PDOS for O 2p, Ti 3d t$_{2g}$ and Ti 3d e$_g$ orbitals. The Fermi level is marked by a dashed line. The PDOS plots are amplified ten times. All spin-up and spin-down channels are marked by the same color for clarity.

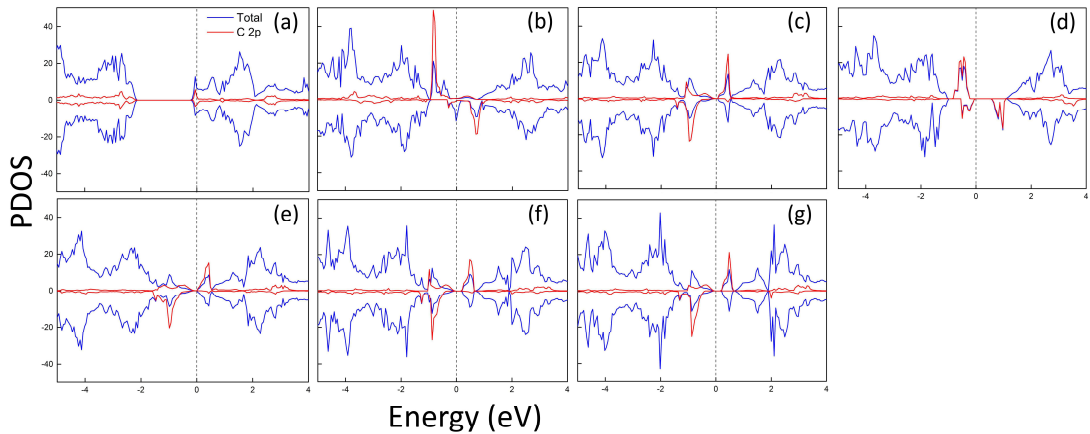

**Fig. 6.** (a)-(g) Calculated DOS for SrTiO$_{2.75}$C$_{0.25}$ with configuration A-G (green) and the PDOS for impurity carbon atoms (red). The Fermi level is marked by a dashed line. The PDOS are amplified ten times. All spin-up and spin-down channels are marked by the same color for clarity.

### 3.3.1 n-type NM metal

From Fig. 6(a), the equal occupancy of spin-up and spin-down states indicates no extra local magnetism in configuration A, which is consistent with our magnetic studies above. On the other hand, it is obvious that both spin-up and spin-down channels cross the Fermi level in the conduction bands compared to bulk $SrTiO_3$, which suggests that configuration A shows n-type nonmagnetic metallic characteristics. Due to the higher energies of the 2p orbitals for foreigner C compared with host O, doped C will introduce an acceptor level and occupy the VBM. The states are somewhat delocalized. The hybridization of C 2p orbitals and Ti 3d orbitals pulls the conduction band minimum (CBM) downwards by approximately 0.22 eV below the Fermi level while compelling the VBM located at 2.08 eV below the Fermi level, which provides a band gap of 1.86 eV. Therefore, the bands originating from C 2p states lead to a widening instead of a narrowing of the band gap.

### 3.3.2 FM/AFM half-metal

The calculated spin-polarized DOS for configuration B are shown in Fig. 6(b), where the uncompensated majority (spin down) and minority (spin up) channels illustrate the existence of magnetic moments. Meanwhile, the majority channels depict a conducting nature, while the minority channels show semiconducting behavior, which gives rise to an FM/AFM half-metal characteristic in configuration B with a second-nearest C-C. As shown in Fig. 6, the DOS for different configurations demonstrates that the semiconductor or insulator feature for Configuration C-G is characterized by a more stable state than the half-metal electronic property of configuration B. Although configuration A is endowed with a metal nature, the induced dimer pair between the two boron dopants makes it the most stable among all configurations. Therefore, configuration B is the most unstable among all configurations, which is in agreement with the observed peculiarity in energy.

It is clear that the introduced C mixed with O 2p states indeed increases the width of the valence band. Moreover, unoccupied C 2p states lead to a downshift of the CBM. Both of these contributions produce a significant narrowing of the band gap.

### 3.3.3 AFM/FM semiconductor

The calculated spin-polarized DOS for configurations C-G are shown in Fig. 6(c)-(g), which all exhibit semiconductor behavior as the C-C distance increases, resulting in AFM or FM semiconductor characteristics in C-G. The introduction of C 2p orbitals hybridized with O 2p states in these configurations leads to a reduction in the band gap. Moreover, additional localized bands are obviously formed near the Fermi level in configurations C-G. Note that even though configurations B and C are endowed with the same C-C distance, they exhibit distinct differences both in terms of the electronic band structures and magnetic peculiarities.

## 3.4 GGA+U

GGA+U calculations are usually applied to deal with strong correlations in localized d- or f-electron systems, and, thus, partially correct the band gap estimation. Therefore, we also employ GGA+U to describe magnetic moments and electronic structures in $SrTiO_{2.75}C_{0.25}$ systems considering the localized d-d interactions. A value of U=5.8 eV is employed to describe Ti 3d orbitals [19], which increases the band gap of pristine $SrTiO_3$ up to 2.57 eV, as shown in Fig. 7(f). The results for the magnetic properties with GGA+U are summarized in Table 3 and Table T2 in the supplemental materials. Table T2 in the supplemental materials shows the averaged contribution of each Sr atom, Ti atom, and O atom to magnetism. In general, the main results from GGA+U are similar to those obtained from GGA calculations. For instance, the relative energy for different configurations from GGA+U follows the same trend as for the GGA calculations. Configuration A shows an NM nature for the GMS accompanied by a remarkable lowest energy, while configuration B is the most unstable state with a relatively high energy among all configurations in both GGA and GGA+U calculations. The GMS for most configurations in GGA+U calculations is also similar to the results obtained from GGA calculations, except for configuration D. The GGA+U calculations predict AFM, while GGA calculations suggest FM/AFM as the GMS of configuration D. Additionally, the electronic structures of configuration A are different in GGA and GGA+U calculations.

Table 3. List of the related energies in NM, FM and AFM states (eV/dopant), $MM_{tot}$, and localized $MM_C$ in units of $\mu_B$, nature of the GMS for different configurations in $SrTiO_{2.75}C_{0.25}$ determined by the GGA+U method. The lowest total energy of configuration A is taken as a reference. The data in bold font highlight the lowest energy (the most stable state) among the NM, FM and AFM alignments for each configuration (note, if the energy difference between FM and AFM is relatively small, e.g., less than typical 25 meV, then both FM and AFM are labeled possible GMSs to take into account the temperature effects and numerical uncertainty).

| Configuration | ΔE(NM/FM/AFM) | $MM_{tot}$(FM/AFM) | $MM_C$(FM/AFM) | GMS |
|---|---|---|---|---|
| A | 0/0/0 | 0/0 | 0, 0/0, 0 | NM |
| B | 2.903/**2.451**/2.490 | 3.810/0.001 | 0.575, 0.855/0.656, -0.656 | FM |
| C | 2.774/2.238/**2.227** | 4.000/-0.004 | 0.721, 0.721/0.711, -0.711 | AFM/FM |
| D | 2.733/2.267/**2.120** | 3.998/-0.002 | 0.713, 0.713/0.721, -0.721 | AFM |
| E | 2.675/2.254/**2.187** | 3.958/-0.003 | 0.722, 0.722/0.692, -0.692 | AFM |
| F | 2.671/2.283/**2.227** | 3.998/0 | 0.722, 0.722/0.693, -0.693 | AFM |
| G | 2.874/2.245/**2.244** | 3.999/-0.002 | 0.717, 0.717/0.716, -0.716 | AFM/FM |

The GGA+U calculated spin-polarized DOS in different configurations is shown in Fig. 7(a)-(g). Configuration B exhibits a half-metal peculiarity, while C-G is characterized by semiconductor behavior, which is the same as that observed in GGA calculations. In addition, more obvious isolated bands are formed above the O 2p orbitals in all configurations under the GGA+U calculations, which can be ascribed to the higher 2p orbitals of carbon compared to the

O 2p orbitals. Compared with the GGA calculation, the dominant difference is that configuration A shows a semiconductor feature rather than a metal nature, which is attributed to the more localized Ti orbitals and weaker Ti-C hybridization under the GGA+U calculation.

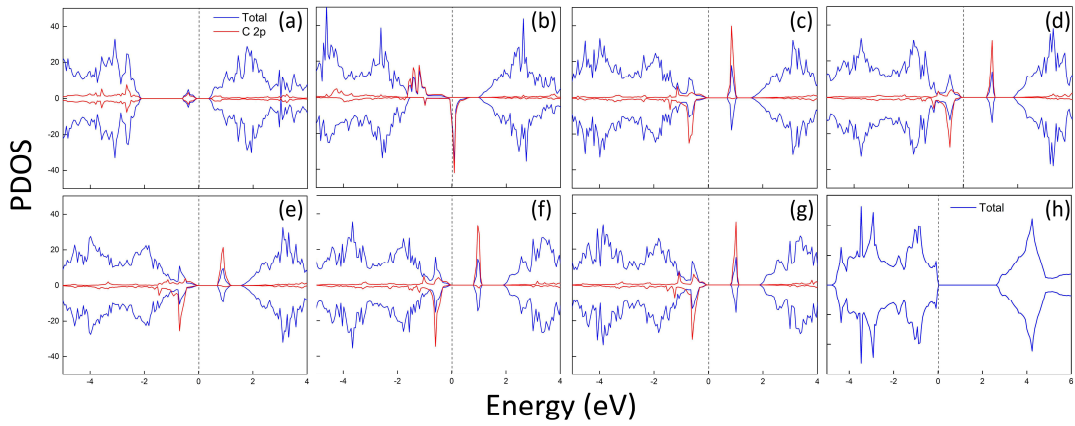

**Fig. 7**(a)-(g): GGA+U calculated DOS for SrTiO$_{2.75}$C$_{0.25}$ with configurations A-G (green) and the PDOS for impurity carbon atoms (red). (h) Total DOS for pristine SrTiO$_3$. The Fermi level is marked by a dashed line. The PDOS are amplified for clarity. All spin-up and spin-down channels are marked by the same color.

As shown in Table 1 and Table 3, GGA and GGA+U calculation results, respectively, show obvious magnetic distinctions in configuration D, i.e., an FM/AFM state versus AFM state in combination D, indicating that the GMS is U affected. To further examine the effects of strongly correlated effects, we perform additional calculations for the energy difference between AFM and FM states in configuration D as a function of different U values and take configuration B for comparison, as shown in Fig. 8. We employ the formula ΔE=$E_{AFM}$-$E_{FM}$ to illustrate the evolution of GMS, where $E_{AFM}$ and $E_{FM}$ represent the energy in the AFM state and FM state, respectively. Within numerical uncertainty, the energy difference ΔE in configuration B as a function of U is rather small, and it tends to be a constant value above 0, indicating an FM/AFM for the GMS. However, for configuration D, a clear crossover is observed from FM to AFM GMS at approximately U=1.6 eV, implying a more stable AFM characteristic in configuration D with strongly correlated effects taken into account.

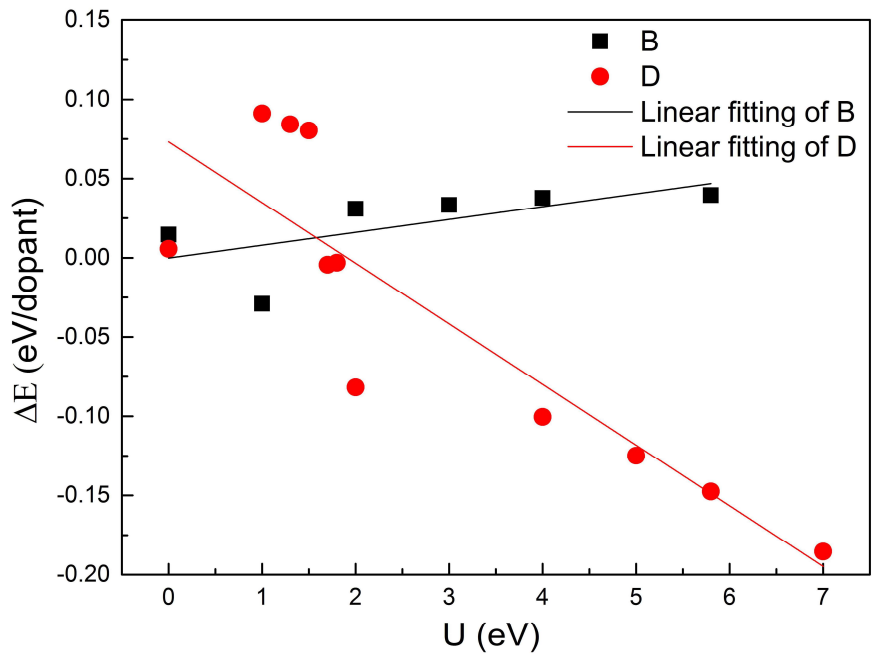

Fig. 8. Energy difference between AFM and FM states in configurations B and D versus different U.S.

## 3.5 Conclusions

The magnetic and electronic properties of different $SrTiO_{2.75}C_{0.25}$ configurations were explored using first-principles calculations with GGA and GGA+U. Our calculation results show that the GMS undergoes a change in magnetic states as the C-C distance increases, *i.e.*, it starts with no noticeable magnetic moments at the shortest C-C distance, followed by a rather small total magnetic moment with unequal carbon contributions on magnetic moments, and, finally, leads to an equal contribution of approximately 0.7 $\mu_B$ of magnetic moment per carbon atom. No noticeable contribution of the magnetic moment is related to Sr and Ti atoms, while a relatively small localized magnetic moment is resolved at O atoms. In both GGA and GGA+U calculations, the ground state exhibits NM features when two carbon atoms are nearest neighbors, which mainly originates from the stronger C-C and weaker C-Ti hybridizations with the shortest C-C distance and leads to equal spin-up and spin-down channels. In addition, both calculations demonstrate that two carbon atoms tend to form a dimer pair with a remarkable displacement of approximately 1.5 Å, while two doped carbons aligned linearly along the C-Ti-C axis are endowed with relatively high energies. The calculation results also show the bandgap modifications when the C-C dopant distance is changed. Our work shows the tunable magnetic and electronic properties of C-doped $SrTiO_3$, which is expected to promote its applications in a variety of fields. In addition, we note the possible doping site of Ti and other sites substituted by C and will investigate these in our future work.

**Acknowledgment**

This project was funded by the National Natural Science Foundation of China (Grant No. 11704317) and supported by the Xiamen University Malaysia Research Fund (Grant No: XMUMRF/2019-C3/IORI/0001).